\documentstyle [12pt] {article}

\catcode`@=12
\begin{document}
\vspace{0.5in}
\oddsidemargin -.375in
\newcount\sectionnumber
\sectionnumber=0
\def\be{\begin{equation}}
\def\ee{\end{equation}}
\begin{flushright} UH-511-809-94\\November 1994\
\end{flushright}
\vspace {.5in}
\begin{center}
{\Large\bf Form factors for Charmed Baryon Decays at order 1/m$_{\bf c}$ in
the HQET
 limit \\}
\vspace{.5in}
{\bf Alakabha Datta \\}
\vspace{.1in}
 {\it
Physics Department, University of Hawaii at Manoa, 2505 Correa
Road, Honolulu, HI 96822, USA.}\\
\vskip .5in
\end{center}

\vskip .1in
\begin{abstract}

We study the vector and axial vector form factors in the heavy
 to light transitions in charmed baryons decays
using HQET.
We calculate the ${1}/{m_c}$ corrections to the form factors and show that
these corrections can be significant. We also consider applications to
semi-leptonic and hadronic decays of charmed baryons.

\end{abstract}
\vskip .25in

With more data on charmed baryons becoming available theoretical studies of
charmed baryons have become important. For transitions of a charmed baryon
into an uncharmed baryon
a total of six vector and axial vector
 form
factors  completely specify the hadronic part of
the matrix element in semi-leptonic charmed baryon decays and the
factorizable part in the two body hadronic decays of charmed baryons. There
exist several calculations in the literature on charmed baryon decays where
these form factors are calculated in specific models \cite{tc}. However,  it is
   interesting to study these form factors in the limit that the charmed
quark is treated as heavy. In the heavy quark limit the spin symmetry of the
heavy quark results in relations among the form factors. In particular,  the
spin symmetry is very effective in reducing the number of independent form
factors for the lowest lying charmed baryons
 because the light degrees of freedom are in a configuration of spin
zero. For heavy to light transitions of the form
 $ c\rightarrow q $, where q is one of the light u,d or s quarks it
 is known that  the six form factors can be expressed in
 terms of only two independent form factors
 in the limit of $ m_c\rightarrow \infty$ \cite{mr}.
 In this work we try to estimate
the ${1}/{m_c}$ corrections to these form factors.

In HQET the Q.C.D Lagrangian for the heavy quark is expanded in inverse
powers of the heavy quark mass. In the limit $ m_{Q}\rightarrow\infty $ the
heavy quark field $ Q(x) $ is replaced by $h_{v}(x) $ \cite{ge} :
\begin{equation}
h_{v}(x)  =  e^{im_{Q}v.x}P_{+}Q(x) \\
\end{equation}
where $ P_{+} = {1+\rlap/v\over2} $ is the positive energy projection
operator. The effective Lagrangian is written as
\begin{equation}
L_{HQET} = \bar h_{v}\,iv\!\cdot\!D\,h_{v}
\end{equation}
where $D^\alpha = \partial^\alpha - ig_{s}t^{a} A^\alpha_a$ is the
gauge covariant derivative.  Corrections to the above
Lagrangian come from higher dimensional operators suppressed
by inverse powers of $m_Q$.  Including ${1}/{m_Q}$
corrections we write \cite{ge,el}
\begin{eqnarray}
L &=& L + \delta L/2{m_{Q}}  \\ \nonumber
\delta L & = & \bar{h} \ (iD)^2 \ h + \frac{g_s}{2} \bar{h} \
\sigma_{\alpha \beta} G^{\alpha \beta} \ h
\end{eqnarray}
where $G^{\alpha \beta} = [iD^\alpha, iD^\beta] = ig_{s}t^{a}G_a^{\alpha
\beta}$ is the gluon field strength.
The equation of motion for the heavy quark is
\begin{equation}
v.Dh_v = 0
\end{equation}

One also has to  expand the currents that mediate the weak
decays of hadrons.  In our case we are interested in
currents of the form $\bar{q} \ \Gamma \ Q$.  At the tree level
the expansion of the current in the HQET gives
\begin{equation}
\bar{q} \ \Gamma \ Q \rightarrow \bar{q} \ \Gamma \ h +
\frac{1}{2m_Q} \bar{q} \ \Gamma \ i  \rlap/D \  h + \cdot
\cdot \cdot
\end{equation}
where $\Gamma$ is any arbitrary Dirac structure.

The vector and axial form factors in the weak decays of
charmed baryons are conventionally parametrized in terms of
6 form factors $f_i$ and $g_i$ defined by
\begin{eqnarray}
 \left < B'(p',s')
\mid \bar{q} \ \gamma^\mu \ Q \mid B_c (p, s)
\right >
\ = \ \bar{u}_{B'} (p', s')
\left [f_1 \gamma^\mu - i \frac{f_2}{m_{B_c}}\sigma^{\mu\nu}
 q_\nu + \frac{f_3}{m_{B_c}} q^\mu \right ] u{_B{_c}} (p, s)
\nonumber \\
 \left < B'(p', s')
\mid \bar{q} \ \gamma^\mu \gamma^5 \ Q \mid B_c (p,s)
\right >
 =  \bar{u}_{B'} (p', s')
\left [g_1 \gamma^\mu - i \frac{g_2}{m_{B_c}}\sigma^{\mu\nu}
 q_\nu + \frac{g_3}{m_{B_c}} q^\mu \right ] \gamma^5 u{_B{_c}} (p, s)
\end{eqnarray}
where $q^\mu = p^\mu-p'^\mu$ is the four momentum transfer.
In our case it is convenient to write a different
parametrization of form factors.
\begin{eqnarray*}
\left < B' (p', s') \mid \bar{q} \ \gamma^\mu \ Q \mid B_c (p,
s) \right >
 =
\bar{u}_{B'} (p', s') \left [F_1 \gamma^\mu + F_2 v^\mu +
\frac{F_3}{m_{B'}} p'^\mu \right ] u_{B_{c}} (p,s)
\end{eqnarray*}
\begin{eqnarray}
\left < B' (p', s') \mid \bar{q} \ \gamma^\mu \gamma^5 \ Q\mid B_c
(p,s) \right >
\ = \ \bar{u}_{B'} (p', s') \left [ G_1 \gamma^\mu + G_2
v^\mu + \frac{G_3}{m_{B'}} p'^\mu \right ] \gamma^5 u_{B_c} (v, s)
\end{eqnarray}
where $v^\mu$ is the velocity of the heavy baryon.
The two sets of form factors are related by
\begin{eqnarray}
f_1 & = & F_1 + (m_{B_c} + m_{B'})
    \left [ \frac{F_2}{2m_{B_c}} + \frac{F_3}{2m_{B'}} \right ]
\nonumber \\
\frac{f_2}{m_{B_c}} & = & - \frac{F_2}{2m_{B_c}} - \frac{F_3}{2m_{B'}}
\nonumber \\
\frac{f_3}{m_{B_c}} &=& \frac{F_2}{2m_{B_c}} - \frac{F_3}{2m_{B'}}
\nonumber \\
g_1 &=& G_1 - (m_{B_c} - m_{B'}) \left [
\frac{G_2}{2m_{B_{c}}} + \frac{G_3}{2m_{B'}} \right ]
\nonumber \\
\frac{g_2}{m_{B_c}} & = & - \frac{G_2}{2m_{B_c}} - \frac{G_3}{2m_{B'}}
\nonumber \\
\frac{g_3}{m_{B_c}} &=& \frac{G_2}{2m_{B_c}} - \frac{G_3}{2m_{B'}}
\end{eqnarray}
In the limit of $m_Q \rightarrow \infty$ the spin symmetry
allows one to write
\begin{equation}
\left < B' \mid \bar{q} \ \Gamma  \ h_v \mid B_c \right > =
\bar{u}_{B'} \
\left [\theta_1 + \rlap/v\ \theta_2 \right ] \Gamma \ U_{B_c} \
\end{equation}
 where $\theta_1, \theta_2$ are Lorentz-invariant and $\Gamma$ is
any arbitrary Dirac structure with
\begin{equation}
\rlap/v U_{B_c}(v, s) = U_{B_c} (v,s)
\end{equation}
  Note the normalization for
$U_{B_c}$, the baryon spinor, is $\bar{U}_{B_c} U_{B_c} = 2M$ where
 $M$ is the effective
mass of the baryon in HQET
and differs from the actual baryon mass by terms
suppressed by second and higher powers of the heavy quark mass. Since we are
interested only in $ 1/m_Q $ corrections these terms can be neglected.
Setting  $\Gamma = \gamma^\mu,
\gamma^\mu \gamma^5$ in eqn.(9) we get the following relations.
\begin{eqnarray*}
F_1^0 &=& \theta_1-\theta_2 \nonumber \\
F_2^0 &=& 2 \theta_2     \nonumber \\
F_3^0 &=& 0  \nonumber \\
\end{eqnarray*}
\begin{eqnarray}
G^0_1 &=& \theta_1 + \theta_2 \nonumber \\
G_2^0 &=& 2 \theta_2 \nonumber \\
G_3^0   &=& 0
\end{eqnarray}
where $ (F^0){_i}, (G^0_{i})$ are the zeroth order form factors in the $ 1/m_Q$
expansion.
Eqn.(11) lead to the following relations between the form
factors \cite{mr}.
\begin{eqnarray}
 G^0_1 &=& F_1^0 + F^0_2 \quad; \quad
G^0_2 = F_2^0 \quad; \quad G^0_3 =0 \quad; \quad F_3^0 = 0  \
\end{eqnarray}
or equivalently,
\begin{eqnarray}
g_1 = f_1 \quad; \quad  g_2 = f_2 \quad; \quad  g_3 \quad = -f_2 \quad; \quad
 f_3=-f_2\
\end{eqnarray}

We next turn to $1/m_c$ corrections to these relations.
The first source of $1/m_c$ corrections comes from the expansion of
the currents in powers of $1/m_c$.  The matrix element of
interest here is
\begin{equation}
\left < B' \mid \bar{q} \ \Gamma  \ i \rlap/D \ h \mid B_c \right >
= \bar{u}_{B'} \phi_{\nu} \Gamma  \gamma^\nu  U_{B_c} \
\end{equation}
where the form of the R.H.S of the above equation follow from
 spin symmetry. The general form of $\phi_\nu$ is
\begin{eqnarray}
\phi_\nu \ = (\phi_{11} v_\nu + \phi_{12}  v'_\nu + \phi_{13}
\gamma_\nu) + {\rlap/v(\phi_{21} v_\nu + \phi_{22}  v'_\nu +
\phi_{23} \gamma_\nu)}
\end{eqnarray}
Using the equation of motion for the heavy quark we have the
following conditions.
\begin{eqnarray}
\left < B' \mid \bar{q} \ i v .D \ h \mid B_c \right >
&=& \bar{u}_{B'}  v . \phi U_{B_c} = 0 \nonumber \\
\left < B' \mid \bar{q}\ \gamma^5 \ i v  . D \ h \mid B_c \right >
&=& \bar{u}_{B'}  v  .  \phi \gamma^5 U_{B_c} = 0
\end{eqnarray}
which results in
\begin{eqnarray}
\phi_{11} + \omega \ \phi_{12} &=& - \phi_{23} = x
\nonumber \\
\phi_{21} + \omega \phi_{22} &=& - \phi_{13} = y
\end{eqnarray}
where $\omega = v.v', v'$ being the velocity of the baryon
$B'$.  Note $x$ and $y$ are defined through the
eqn. (17).
The corrections to the form factor can now be easily written
down from eqn. (15).
\begin{eqnarray}
\delta F_1 &=&\frac{\left[ \left ( \frac{\omega + 1}{\omega} \right )
\left (\phi_{11} + \phi_{21} \right ) - \frac{y}{\omega} -
\left (\frac{2 \omega +1}{\omega} \right ) x\right]}{2m_c}
\nonumber \\
\delta F_2 &=& \frac{2 \left [ \phi_{21} \left ( \frac{\omega -1}
{\omega} \right ) + 2 x + \frac{y}{\omega} \right ]}{2m_c}
\nonumber \\
\delta F_3 &=&\frac{ 2 \left [ -\frac{\phi_{11}}{\omega} -
\frac{\phi_{21}}{\omega} + \frac {(x + y)}{\omega} \right ]}{2m_c}
\nonumber \\
\delta G_1 &=&\frac{ \left[\left ( \frac{1-\omega}{\omega} \right) \left (
\phi_{11}- \phi_{21} \right ) + \frac{2 \omega -1}{\omega}
x + \frac{y}{\omega} \right]}{2m_c} \nonumber \\
\delta G_2 &=&\frac{ 2 \left [ - \phi_{21} \left ( \frac{\omega + 1}
{\omega}\right ) + 2 x + \frac{y}{\omega} \right ]}{2m_c}
\nonumber \\
\delta G_3 &=&\frac {2 \left [ -\frac{\phi_{11}}{\omega} + \frac
{\phi_{21}}{\omega} +  \frac{(x-y)}{\omega} \right ]}{2m_c}
\end{eqnarray}
Using the equation of motion one can derive two relations between the the
 $\phi$'s. We start with
\begin{eqnarray*}
\left < B' \mid i \partial_\mu  ( \bar{q}\ \Gamma \ h_v)
\mid B_c \right > = \left < B' \mid  [ i ( \partial_\mu +
  i g_s  t_a
 A_\mu^a  ) \bar{q}] \ \Gamma \ h_v
 + \bar{q} \ \Gamma \  [ i ( \partial_\mu - i  g_s t_a  A_\mu^a)
 h_{v}]   \mid B_c \right > \
\end{eqnarray*}
\begin{equation}
 \quad \quad = \ \left [ (M-m_Q) v_\mu - m_{B'} v'_\mu \right ]
\left < B' \mid \bar{q} \ \Gamma \  h_v \mid B_c \right >
\end{equation}
where $M$ is the heavy baryon mass in the $HQET$ limit.  We
can rewrite the above identity as
\begin{eqnarray}
\left < B' \mid  (i D^*_\mu  q  ) \ \Gamma \  h_v
+ \bar{q} \ \Gamma \ (i  D_\mu  h_v) \mid B_c \right >
& = & \left (\bar{\Lambda} v_\mu - m_{B'}  v'_\mu \right )
\left < B' \mid \bar{q} \ \Gamma \ h_v \mid B_c \right >
\end{eqnarray}
where $ \bar{\Lambda}=M-m_c$.
With $\Gamma = \ \gamma^\mu  \stackrel{\wedge}{\Gamma}$ and using the
 equation of
motion for the light quark $q$  one can write
\begin{equation}
\left < B' \mid \bar{q} \ \gamma^{\mu}  \stackrel{\wedge}{\Gamma} i  D_\mu
 \ h_v \ \mid B_c \right > = [\bar{\Lambda} v_\mu - (M_{B'} - m_q)
v'_\mu] \left < B' \mid \bar{q} \ \gamma^\mu \stackrel{\wedge}{\Gamma}
 h_v
\mid B_c \right >
\end{equation}
With $\stackrel{\wedge}{\Gamma} = 1$ one obtains,
\begin{eqnarray}
( \phi_{11}-\phi_{21}) \frac{\omega - 1}{\omega} -
\left ( \frac{2 \omega +1}{\omega} \right ) y - \left( \frac{4
\omega - 1}{\omega}\right ) x \nonumber \\
\ = \ F^0_1 (\bar{\Lambda} - \stackrel{\wedge}{m}_{B\prime}) + F_2^0
(\bar{\Lambda} -
\stackrel{\wedge}{m}_{B\prime}
\omega)
\end{eqnarray}
where $F_1^0, F_2^0$ are the vector form factors in $m_c\rightarrow\infty$
limit and $\stackrel{\wedge}{m}_B=m_{B'}-m_q$,
while with $\stackrel{\wedge}{\Gamma} = \gamma_5$ one obtains,
\begin{eqnarray*}
\left ( \phi_{11} + \phi_{21} \right ) \frac {\omega +
1}{\omega} + \left ( \frac{4 \omega - 1}{\omega}\right )x +
\left ( \frac{2 \omega - 1}{\omega} \right )y
\end{eqnarray*}
\begin{eqnarray}
& = & G^0_1 (\bar{ \Lambda} + \stackrel{\wedge}{m}_ {B'}) - G_2^0 \left
 (\bar{\Lambda} -
\stackrel{\wedge}{m}_{B'} \omega \right )
\end{eqnarray}
where
$G_1^0, G_2^0$ are the axial vector form factors in the $m_c \rightarrow
\infty$ limit.  One can now solve for $\phi_{11}$,
$\phi_{21}$ in terms of $F_1^0, F^0_2, G_1^0, G^0_2, x$ and y

We note that at $\omega=1$
\begin{equation}
 y = -\frac{G_1^0 (w=1) (\bar{\Lambda} - \stackrel{\wedge}{m'}_B)}{3}-x
\end{equation}

We see therefore that at $\omega=1$ the corrections to the form
factor can be expressed in terms of the zeroth order form
factors and $x$.  If we further assume that $x \sim y$ with both $x$ and $y$
being small in the spirit of $1/m_Q$ expansion, then
\begin{equation}
x \sim -\frac{G_1^0 (\bar{\Lambda}- \stackrel{\wedge}{m}_ {B'})}{6}
\end{equation}
 so all the corrections to the form factors can be expressed, at $\omega
=1$, in terms of the zeroth order form factors. ( Note that for heavy-heavy
transition $\bar{\Lambda}\simeq \stackrel{\wedge}{m}_ {B'}$ upto  order $1/m_Q$
and so $x\simeq 0$ which is what is expected. )
The second source of $1/m_Q$ corrections comes from
${\delta L}/{2 m_Q}$ and one has to calculate
\begin{eqnarray*}
\frac{1}{2m_Q} \left < B' \mid i \int \ dy \ T \left \{
\bar{q} \ \Gamma \ {h_v}(0),\ \delta L(y) \right \} \mid  B_c
\right >_{HQET}
\end{eqnarray*}
Using spin symmetry we can write the corrections from the
kinetic energy operator as
\begin{eqnarray}
\frac{1}{2m_Q} \left < B' \mid i \int \ dy \ T \left \{
\bar{q} \ \Gamma \ {h_v}(0), \ \bar{h} \ (iD)^2 \ h(y) \right \} \mid  B_c
\right
>_{HQET} \\
\ = \ \frac{A (\omega)}{2m_Q} \left < B' \mid \bar{q} \ \Gamma \
h_v (0) \mid B_c \right >
\end{eqnarray}
Clearly this correction only renormalizes the zeroth order
form factors.  We also note that we can replace the zeroth order form
factors in eqns. (22) and (23)
 by the  renormalized form factors because the
corrections are of the order ${1}/{(2 m_Q)^2}$
which we neglect.

The corrections from the chromomagnetic operator can be
written as, using spin symmetry,
\begin{eqnarray}
\frac{1}{2m_Q} \left < B' \mid i \int \ dy \ T \left \{
\bar{q} \ \Gamma \ h_v(0), \frac{g_s}{2} \ \bar{h} \ \sigma_{\alpha
\beta}  G^{\alpha \beta} \ h (y) \right \} \mid  B_c \right
>_{HQET} \nonumber \\
\ = \ \bar{u}_{B'}\ \chi_{\alpha \beta} \ \Gamma \left ( \frac{1
+ \rlap/v}{2} \right ) s^{\alpha \beta} \ U_{B_c} \ \ \mbox{where}
\end{eqnarray}
\begin{eqnarray*}
\chi_{\alpha \beta} = (\chi_{11} + \rlap/v\ \chi_{12})
\frac{(\gamma_\alpha \gamma_\beta - \gamma_\beta
\gamma_\alpha)}{4} + (\chi_{21} + \rlap/v\ \chi_{22})
\frac{(\gamma_\alpha v'_\beta - \gamma_\beta v'_\alpha)}{4}
\\
\mbox{where} \quad s^{\alpha \beta} \ = \
\frac{\gamma^{\alpha} \gamma^{\beta}- \gamma^{\beta} \gamma
^{\alpha}}{4}\\
\end{eqnarray*}
The corrections to the form factors are then
\begin{eqnarray*}
\Delta F_1 & = &\frac{ \left [ (\chi_{11} - \chi_{12}\right ]}{2} \\
\Delta F_2 & = & \left [-2 \chi_{11}-  \chi_{12}  +
\frac{\chi_{21}}{2} + (2 \omega -1)\frac{\chi_{22}}{2} \right ] \\
\Delta F_3 & = &\frac{ [- (\chi_{21}\ + \chi_{22})]}{2} \\
\Delta G_1 & = & \frac{\left   [(\chi_{11} + \chi_{12})  \right ]}{2} \\
\Delta G_2 & = & \left [ 2 \chi_{11} - \chi_{12} - \frac{\chi_{21}}{2}
 - (2 \omega
+1)
\frac{\chi_{22}}{2}  \right ] \\
\Delta G_3 & = &\frac{\left[  (\chi_{22}-\chi_{21})\right]}{2}
\end{eqnarray*}
 Therefore we see that for corrections from the chromagnetic term
 there
are
four new  matrix elements that have to be calculated to estimate the $1/m_Q$
corrections.

It is straightforward to include radiative corrections in
our analysis.  For instance the relations between form
factors in the $m_Q \rightarrow \infty$ limit get modified in the
presence of radiative corrections to \cite {ne}
\begin{eqnarray}
G_1 & = & \left [ \frac{C_1 (\mu)}{C_1 (\mu) + C_2 (\mu)}
\right ] [F_1 + F_2] \\ \nonumber
G_2 & = & \frac{C_1 (\mu)}{C_1 (\mu) + C_2 (\mu)} F_2 -
\frac{C_2 (\mu)}{C_1 (\mu)} F_1
\left [ 1 + \frac{C_1 (\mu)}{C_1 (\mu) + C_2 (\mu)} \right ] \\
\nonumber
G_3 &=& F_3 \ = 0
\end{eqnarray}
where $C_1 (\mu), C_2 (\mu)$ are the Wilson's co-efficients
that occur in the short-distance expansion of the currents.
At the ${1}/{m_Q}$ level there are extra operators that
arise namely \cite{ne,al}
\begin{eqnarray}
O_1 & = & \bar{q} \ v^\mu \ i \rlap/D \ h_v \quad ; \quad  O_2 =
\bar{q} \ i D^\mu \ h_v \quad ; \quad O_3 = \bar{q} \  (-
iv.\stackrel{\leftarrow}{D}) \ \gamma^\mu \ h_v \nonumber \\
O_4 & = & \bar{q} \ (-iv.\stackrel{\leftarrow}{D}) \ v^\mu \ h_v \quad ;
 \quad O_5 =
\bar{q} \ (-i \stackrel{\leftarrow}{D}^\mu) \ h_v \\
\nonumber
O_6 & = & m_q \ \bar{q} \ \gamma^\mu \ h_v \quad ;
 \quad O_7 = m_q \ \bar{q} \ v^\mu \ h_v
\end{eqnarray}
and the chromomagnetic operates also gets renormalised.
It is straightforward to calculate the matrix elements of
operators $O_1 - O_5$ using equations (15) and (20).  The matrix
elements of the operators $O_6$ and $O_7$ will just renormalize
 the zeroth order form factors.  Radiative corrections are
typically $\sim$ 5\% and  we neglect these corrections in
our calculations.

To use the results of the above section in phenomenology one has to make
estimates of the six new quantities $x$, $y$, $\chi_{11}$, $\chi_{12}
$, $\chi_{21}$, and $\chi_{22}$. These
quantities are uncalculable in HQET and one has
to use other techniques (Q.C.D sum rules for example) to calculate them.
 In this
work we do not address the problem of calculating these quantities but we try
to get some estimates of the quantities using some reasonable assumptions.

Starting with the corrections coming from the expansion of the current, a
reasonable assumption is the one made in eqn.(25). Even for $\omega$
near $1$ eqn.(25) can be used to estimate $x$ provided $x$ is a
 slowly varying function of $\omega$.

For $\omega$ near 1  and making the approximation
$x (\omega) \sim y (\omega) \sim x (\omega =1) \sim y (\omega =
1)$ equations (22) and (23) reduce to
\begin{eqnarray*}
(\phi_{11} - \phi_{21}) \frac{\omega - 1}{\omega} & = &
F_1^0 (\bar{\Lambda} - \stackrel{\wedge}{m}_{B'}) + F_2^0 (\bar{\Lambda}
- \stackrel{\wedge}{m}_{B'} \omega) - {G_1^0 (\bar{\Lambda} -
\stackrel{\wedge}{m}_{B'})} \\
& = & F_2^0 (1 -\omega) \stackrel{\wedge}{m}_{B'}  \\
(\phi_{11} + \phi_{21}) \frac{\omega + 1}{\omega} & = &
G_1^0 (\bar{\Lambda} + \stackrel{\wedge}{m}_{B'}) - G_2^0 (\bar{\Lambda}
- \stackrel{\wedge}{m}_{B'} \omega) \\
& + & \left [ 1 - \frac{1}{3 \omega} \right ] G_1^0 (\bar{\Lambda} -
\stackrel{\wedge}{m}_{B'})
\end{eqnarray*}
For the factorized two body hadronic decays of charmed baryons taking for
example $\Lambda_c \rightarrow \Lambda \pi$ and $\Lambda_c
\rightarrow \Lambda \rho$ we find $\omega \sim 1.25$ and
$\omega \sim 1.12$ respectively and so  the
 above approximation is better for vector boson decays.
We see that as far as corrections from
expansion of currents are concerned the largest correction is
$\sim {G_1^0 (\bar{\Lambda} + \stackrel{\wedge}{m}_ {B'})}/{2m_c}
\sim .6 G_1^0$  which is quite significant.

For the corrections from the chromomagnetic operator even at
$\omega = 1$ there are four unknown functions $\chi_{11} ,
\chi_{12} , \chi_{21}$ and $ \chi_{22}$.
We do not have estimates of these functions but the
contribution of chromomagnetic operators have been calculated
to be small in the case of mesons \cite{mn} and also in the case
of heavy-heavy transitions in baryons the matrix element of the chromomagnetic
operator vanishes in first order in ${1}/{m_Q}$
expansion \cite{ggw}.  In the case of heavy to heavy transitions at
$\omega=1$ there is a normalization condition which comes
from the conservation of the flavour conserving vector
current in the limit of equal hadron masses \cite{lu,bf}.
 This absolute normalisation
relation involving the form factors can be used to find relations among matrix
elements, that arise as corrections to the form factors in
 the $1/m_Q$ expansion, at $\omega=1$.
In the case of heavy-light transition however, we do not
have such a normalization condition. There have been attempts to use HQET
 in $c
\rightarrow s$ transitions treating both c  and s quark as heavy
\cite{kk,kk1}.
The basic assumption involved in such an analysis is the following ;
in HQET on the scale of the heavy quark mass the light degrees of
freedom have small momentum spread about their central equal
velocity value.  For strange baryon or meson this is not
true, however, it is possible that the smearing of the momentum of the light
degrees
 averages out
effectively.
In the limit of equal hadron masses we would then have the
normalization condition at $\omega =1 $
\begin{eqnarray}
F_1 + F_2 + F_3 & = & 1   \quad  \mbox{which implies}  \\
F_1^0 + F_2^0 & = & 1
\end{eqnarray}
we therefore get the condition that
\begin{eqnarray}
\delta F_1 + \delta F_2 + \delta F_3 = 0 \ \ \mbox{in the limit of
equal hadron masses }
\end{eqnarray}
In this work we do not assume the validity of an $1/m_s$ expansion but we
make the  assumption that that eqn.(33) is valid upto to the order we
are working in even for unequal hadron masses or at most the R.H.S of
eqn.(33) $\sim x/2m_c$ for unequal hadron masses.
 This is indeed the case in heavy to heavy transitons
 where for example both eqns. (32) and (33) are true
for $\Lambda_b\rightarrow\Lambda_c$ upto $1/m^2_Q$
for unequal hadron masses and is a consequence of Luke's
theorem \cite{bf}. In our case it is unnecessary to use eqn.(32) to estimate
the
 chromomagnetic
corrections. Further making the assumption
 $\chi_{11}\sim\chi_{12} \sim \chi_{21}\sim\chi_{22}$
 one
obtains using
$\delta F_1 + \delta F_2 + \delta F_3 = 0$
\begin{eqnarray*}
\begin{array}{c}
 \chi_{11} = \chi_{12} =\chi_{21} = \chi_{22} = \frac{x}{m_c}
\end{array}
\end{eqnarray*}
where we have set all the $\chi$'s equal to get an estimate of the
chromomagnetic corrections.
Hence the corrections to the form factors at $\omega = 1$
can be written as
\begin{eqnarray}
\begin{array}{ll}
\Delta F_1 = 0    &   \Delta G_1 = \frac{x}{m_c} \\
\Delta F_2 = -\frac{2x}{m_c}    &   \Delta G_2 = -\frac{x}{m_c}\\
\Delta F_3 = \frac{x}{m_c}         &   \Delta G_3 = 0
\end{array}
\end{eqnarray}
We see that the corrections from the chromomagnetic operator is rather small
since $ {x}/{2m_c}\sim $ few percent of $G^0_{1}$. Note that if we had
used $\delta F_1 + \delta F_2 + \delta F_3 \sim {x}/{2m_c}$ the order of
the corrections to the form factors would still
 be same as in the above equation.
  Therefore we have
been able to express the $1/m_c$ corrections to the form factors at and
near $\omega=1$ in terms of the zeroth order form factors ${F_{1}}^0$ and
${F_{2}}^0$. Away from the zero recoil point $\omega=1$ or $q^2=q^2_{max}$
point one may chose a dipole form for the $q^2$ dependence and write a
generic form factor $F$ as \cite{kk1}
\begin{equation}
F(q^2)=\frac{F(q^2_{max})}{(1-\frac{q^2}{{m_{FF}}^2})^2}
{(1-\frac{q^2_{max}}{{m_{FF}}^2})}^2
\end{equation}
 where $m_{FF}$ is the appropriate pole mass. We see therefore that
 we can study
heavy to light transitions in charmed baryons in terms of only two form
factors evaluated at the zero recoil point. Note that in the limit
$m_c\rightarrow\infty$ there are also two independent form factors. However
because of $1/m_c$ corrections the relations in eqns.(12) and(13)
 are no longer valid. As an application of our results we can study the
various asymmetries in exclusive semi-leptonic decays of $\Lambda_c$. The
asymmetries are expressed in terms of helicity amplitudes which in turn can
be expressed in terms of the form factors \cite{kk1}. Since all the form
factors are expressible, including $1/m_c$ corrections in terms of
  $F^0_1$ and $ F^0_2$  all the asymmetries can be expressed as
functions of $F^0_2/F^0_1$. We can write the form factors in eqn.(6) as
\begin{eqnarray}
\frac{f_1}{F^0_1} & = & 1+a + (m_{B_c} + m_{B'})
    \left [ \frac{r+b/3}{2m_{B_c}} - \frac{a+b/3}{2m_{B'}} \right ]
\nonumber \\
\frac{f_2}{F^0_1 m_{B_c}} & = & - \frac{r+b/3}{2m_{B_c}} + \frac{a+b/3}{2m_B'}
\nonumber \\
\frac{f_3}{F^0_1 m_{B_c}} &=& \frac{r+b/3}{2m_{B_c}} + \frac{a+b/3}{2m_B'}
\nonumber \\
\frac{g_1}{F^0_1} &=& 1+r+\frac{2b}{3} - (m_{B_c} - m_B') \left [
\frac{r-a - \rho b/3}{2m_{B_{c}}} + \frac{\rho b/3}{2m_B'} \right ]
\nonumber \\
\frac{g_2}{F^0_1 m_{B_c}} & = & - \frac{r-a-\rho b/3}{2m_{B_{c}}} -
\frac{\rho b/3}{2m_B'}
\nonumber \\
\frac{g_3}{F^0_1 m_{B_c}} &=& \frac{r-a - \rho b/3}{2m_{B_c}} -
\frac{\rho b/3}
{2m_B'}
\end{eqnarray}
where
\begin{eqnarray}
z_1 = (\bar{ \Lambda} + \stackrel{\wedge}{m}_ {B'}) \quad ; \quad
z_2 = (\bar{ \Lambda} - \stackrel{\wedge}{m}_ {B'})
\quad ;\quad r = F^0_2/F^0_1 \quad ; \quad
\rho = -\frac{6r}{1+r}{\frac{\stackrel{\wedge}{m}_ {B'}}{z_2}}
\nonumber\\
a = \frac{(z_1+\frac{4}{3}z_2)+r(z_1+\frac{1}{3}z_2)}{2m_c} \quad  ;
\quad  b = -\frac{(1+r)}{2m_c}z_2
\end{eqnarray}
In the above equation we have included corrections from the chromomagnetic
operator also though the maximun correction from this source is $ 2b/3\sim
-0.04G^0_1$ and so may be neglected
The helicity amplitudes are given by
\begin{eqnarray}
H_{\lambda_2,\lambda_{W}}  =  H^V_{\lambda_2,\lambda_{W}}+
 H^A_{\lambda_2,\lambda_{W}}\nonumber\\
H^{V,(A)}_{-\lambda_2,-\lambda_{W}} =+ (-) H^{V,(A)}_{-\lambda_2,-\lambda_{W}}
\end{eqnarray}
where $\lambda_2,\lambda_{W}$ are the polarizations of the daughter baryon and
the W-boson. In terms of the form factors the helicity amplitudes are given by
\begin{eqnarray}
H^{V}_{1/2,0} = a_{-}\left [(M_1+M_2)f_1+\frac{q^2}{m_{B_c}}f_2\right ]V(q^2)
\nonumber\\
H^{A}_{1/2,0} = a_{+}\left [-(M_1-M_2)g_1+\frac{q^2}{m_{B_c}}g_2\right ]A(q^2)
\nonumber\\
H^{V}_{1/2,1} = \sqrt{2Q_{-}}\left [ -f_1 - \frac{(M_1+M_2)}{m_{B_{c}}}
f_2 \right ]V(q^2)
\nonumber\\
H^{A}_{1/2,1} = \sqrt{2Q_{+}}\left [ g_1 - \frac{(M_1-M_2)}{m_{B_c}}g_2
\right ]A(q^2)
\end{eqnarray}
where
\begin{eqnarray}
Q_{\pm} = {(M_{1} \pm  M_2)}^2 - q^2 \quad ; \quad a_{\pm} =
\sqrt{\frac{Q_{\pm}}{q^2}}\nonumber\\
V(q^2) = \frac{(1-\frac{q^2_{max}}{{m^V_{FF}}^2})^2}
{(1-\frac{q^2}{{m^V_{FF}}^2})^2} \quad ; \quad
A(q^2) = \frac{(1-\frac{q^2_{max}}{{m^A_{FF}}^2})^2}
{(1-\frac{q^2}{{m^A_{FF}}^2})^2}
\end{eqnarray}
where $M_1,M_2$
are the parent and daughter baryon masses and
 $m^{V,A}_{FF}$ are the appropriate pole masses. The asymmetries can
now be calculated using the expressions given below \cite {kk1}
  and will depend only
on the ratio $F^0_2/F^0_1 $.
\begin{eqnarray}
\alpha &=& \frac{|H_{1/2 \ 1}|^2 - |H_{-1/2 \ -1}|^2 + |H_{1/2 \ 0}|^2
- |H_{-1/2 \ 0}|^2}{|H_{1/2 \ 1}|^2 + |H_{-1/2 \ -1}|^2 + |H_{1/2 \ 0}|^2
+ |H_{-1/2 \ 0}|^2} \nonumber\\
\alpha^{\prime} &=& \frac{|H_{1/2 \ 1}|^2 - |H_{-1/2 \ -1}|^2 }
{|H_{1/2 \ 1}|^2 + |H_{-1/2 \ -1}|^2 +2 \ (|H_{1/2 \ 0}|^2
+ |H_{-1/2 \ 0}|^2)}\nonumber\\
\alpha^{\prime \prime} &=& \frac
{|H_{1/2 \ 1}|^2 + |H_{-1/2 \ -1}|^2 - 2 \ (|H_{1/2 \ 0}|^2
+ |H_{-1/2 \ 0}|^2)}
{|H_{1/2 \ 1}|^2 + |H_{-1/2 \ -1}|^2 +2 \ (|H_{1/2 \ 0}|^2
+ |H_{-1/2 \ 0}|^2)} \nonumber\\
\gamma &=& \frac{2 \ Re  (H_{-1/2 \ 0}H^*_{1/2 \ 1} +
 H_{1/2 \ 0}H^*_{-1/2 \ -1})}
{|H_{1/2 \ 1}|^2 + |H_{-1/2 \ -1}|^2 + |H_{1/2 \ 0}|^2
+ |H_{-1/2 \ 0}|^2}\
\end{eqnarray}
 for unpolarized $\Lambda_c$ while for polarized $\Lambda_c$ one has the
following asymmetries \cite {kk1}
\begin{eqnarray}
\alpha_{P} &=& \frac{|H_{1/2 \ 1}|^2 - |H_{-1/2 \ -1}|^2 - |H_{1/2 \ 0}|^2
+ |H_{-1/2 \ 0}|^2}{|H_{1/2 \ 1}|^2 + |H_{-1/2 \ -1}|^2 + |H_{1/2 \ 0}|^2
+ |H_{-1/2 \ 0}|^2} \nonumber\\
\gamma_{P} &=& \frac{2 \ Re (H_{1/2 \ 0}H^*_{-1/2 \ 0})}
{|H_{1/2 \ 1}|^2 + |H_{-1/2 \ -1}|^2 + |H_{1/2 \ 0}|^2
+ |H_{-1/2 \ 0}|^2}\
\end{eqnarray}
 We can also calculate the asymmetries in
hadronic decays of the charmed baryons as a function of $F^0_2/F^0_1 $
which will be given elsewhere.
  A fit to the data on
 the semi-leptonic decay of $ \Lambda_c $
  performed by taking into account the $1/m_c$
 corrections  would result in a value for the ratio $F^0_2/F^0_1$. To
calculate the absolute decay rates one needs $F^0_1$ and $F^0_2$
seperately. This may be fixed from the  measurement of absolute
 decay rates in semi-leptonic decays or
hadronic decays of the charmed baryon though for hadronic decays one has to
contend with
extra theoretical uncertainties.

Summarizing we have calculated the ${1}/{m_c}$
corrections  to the weak hadronic form factors for charmed
baryon decays.  Using certain assumptions about matrix elements we
can estimate these ${1}/{m_c}$ corrections at or near
$\omega \sim 1$ in terms of two form factors $F^0_1, F^0_2$. With a dipole
form for the form factors one can extrapolate these form factors to
arbitrary $q^2$.
One can now use these results to study the weak decays of
charmed baryons involving the transition of a heavy quark into a light quark.

{\bf Acknowledgement } :
I would like to thank Professor Sandip Pakvasa for useful discussions. This
 work was supported in part by US
D.O.E grant \# DE-FG 03-94ER40833.

\end{document}